\documentstyle[12pt,epsfig]{article}

\newlength{\dinwidth}
\newlength{\dinmargin}
\begin{document}
\rightline{SNUTP~95-010}
{}~~\\
\vspace{1cm}
\begin{center}
\begin{Large}
\begin{bf}
%
%
\renewcommand\thefootnote{\fnsymbol{footnote}}
INVESTIGATION OF $WW\gamma$ COUPLINGS
\footnote{Invited talk given at the Korea-Japan Joint Symposium 94,
Dec. 13-15, 1994, Seoul, Korea.
Proceedings to be published by the World Scientific, Singapore,
edited by I. T. Cheon.}
\\
\end{bf}
\end{Large}
\vspace{5mm}
\begin{large}
%
%
\renewcommand\thefootnote{\alph{footnote}}
\setcounter{footnote}{0}
C. S. Kim$^{1,}$\footnote{kim@cskim.yonsei.ac.kr},
Jungil Lee$^{2,}$\footnote{jungil@phyy.snu.ac.kr},
H. S. Song$^{2,}$\footnote{hssong@phyy.snu.ac.kr}\\
\end{large}
%
%
$^{1}$ Department of Physics, Yonsei University, Seoul 120-749, Korea\\
$^{2}$ Center for Theoretical Physics and Department of Physics,\\
Seoul National University, Seoul 151-742, Korea
\vspace{5mm}
\end{center}
\noindent
%
%
\begin{abstract}
We reviewed how measurements of weak boson production
at high energy $ep$ and $e\gamma$ collisions can provide important
information on anomalous $WW\gamma$ couplings. We also considerd the
sinlge muon production through the virtual $W$-decay at the Pohang
Light Source (PLS) facility, and found this process is not adequate
to be detected at the PLS until a large luminosity
($\sim 10^{33}$/sec/cm$^2$)
Free Electron Laser is installed.
\end{abstract}
%
%

\section{Introduction}

Despite impressive experimental confirmation of the correctness
of the Standard Model~(SM), the most direct consequence of the
$SU(2) \times U(1)$ gauge symmetry, the nonabelian self-couplings
of $W,Z$, and photon remains poorly measured to date.
Furthermore, gauge boson coupling strengths are strongly constrained
by gauge invariance, and are sensitive to deviations from the SM. Hence,
experimental bounds on these couplings might shed light on new physics
beyond the SM.

In order to parametrize non-standard effects, it is important to know
what sort of additional couplings
can arise once the restrictions due to gauge
invariance are lifted. As has been previously shown\cite{Hagiwara},
there can be $14$ or more non-standard couplings in the most general
case. To keep the analysis manageable, we restrict ourselves to C,~P
and $U(1)_{\mbox{em}}$ conserving couplings.
This restriction leads to just two
anomalous form factors for the $WW\gamma$ couplings,
traditionally denoted by $\lambda_\gamma$ and
$\kappa_\gamma$
in the $WW\gamma$ sector of the SM, which can be related to the anomalous
electric quadrupole and the anomalous magnetic dipole moment
of the $W$\cite{Aronson}. In the SM at tree level, $\lambda_\gamma=0$ and
$\kappa_\gamma=1$. At present the best experimental limits,
$-3.6<\lambda_\gamma<3.5$ and $-3.5<\kappa_\gamma<5.9$,
are from a recent analysis
of the $W\gamma$ production at s$p\bar{p}$s
by UA(2) collaboration~[3]. While these bounds are compatible with the SM,
they are still too weak to really be considered as a precision test of the SM.
Furthermore, in the absence of beam polarization, it is unlikely that
there will be a significant improvement from the study of $W$ pair production
at LEP-II\cite{Zeppenfeld}.

At future high energy $e^+e^-$ and $e\gamma$ colliders,
probing of the $WW\gamma$ vertex can be performed more precisely
\cite{ee,Yeh}.
One can consider several processes at those colliders.
Among them the process $e+\gamma\rightarrow W+\nu$ has been preferred.
This process has several advantage over the others such as
$e^+e^-\rightarrow W^+W^-$ which also has $WWZ$ vertex.
If we restrict the decay products of $W$ as $\mu+\bar{\nu}_\mu$,
we have a very clean, virtually background-free, events.
There are no final particles detected other than $\mu$ and missing $p_T$
is attributed to the two neutrinos($\nu_e, \bar{\nu}_\mu$).
In view of the detected particle,
we must take into account the process,
$e+\gamma\rightarrow W^*(\rightarrow \mu+\bar{\nu}_\mu)+\nu_e$.

We can also consider photoproduction of a single $W$ boson
at $ep$ colliders.
In $ep$ collision, hadronic jets are
produced due to the subprocess
$\gamma+q\rightarrow W+q^\prime$
and it will
provide a precise test of the structure of the Standard Model
\hspace{0.1cm}$WW\gamma$ vertex.
And the situation there is much cleaner, for
 example, than in $pp$ or $p\bar{p}$ colliders,
where a $W$ and a photon have to be
identified in the final state\cite{Alitti}.

Theoretical studies of the $WW\gamma$ vertex at $ep$ colliders
have been performed\cite{BauerZ,Kim,Jungil}.
The measurement of $\kappa_\gamma$ at $ep$ colliders using the shape of the
$p_{_T}$ distribution of $W$ production at large $p_{_T}$ has been
previously investigated in\cite{BauerZ}. However, this method
suffers from the disadvantage of being sensitive to uncalculated
higher-order QCD corrections, uncertainties in the parton distribution
of the photon, experimental systematic uncertainties, etc~\cite{Bawa}. We have
previously found\cite{Kim,Jungil}
that a measurement of the anomalous coupling
in the $WW\gamma$ vertex at $ep$ colliders can best be achieved by
considering the ratio of the $W$ and $Z$ production cross sections.
The advantage of using a cross section ratio is that uncertainties
from the luminosity, structure functions, higher-order corrections,
QCD scale, etc. tend to cancel\cite{Kim}.
Recently, we investigated the possibility of measuring
both $\kappa_\gamma$ and $\lambda_\gamma$ at the same time by considering
the total cross sections of massive gauge bosons $W$ and $Z$
at $ep$ colliders\cite{Jungil}.

In this paper, we  study the anomalous $WW\gamma$ vertex
by using $ep$ and $e\gamma$ colliders.
In section 2, the most general $WWV(=\gamma, Z)$
vertex, which is C and P even, is reviewed.
We discuss possible processes which produce
single $W$ in $e\gamma$ and $ep$ collision.
In $ep$ collision, we review the techniques
to deal with the single $W$ production process in detail.
We show that the ratio of $W$ and $Z$ production cross sections
is particularly well suited to an experimental determination of the
anomalous $WW\gamma$ coupling parameters
$\kappa_\gamma$ and $\lambda_\gamma$, being relatively insensitive to
uncertainties in the theoretical and experimental parameters.
Section 3 is devoted to the discussion of the process
$\gamma + e \rightarrow \mu+\bar{\nu}_\mu+\nu_e$.
In view of the decay products, we emphasize the advantages of the process
and derive the simplified  squared amplitude
of this process, where $W$ is virtual, in a factorized form.
Discussion is given in section 4.
\section{Reviews on the Anomalous Tri-boson Vertex
in Single $W$ Production}

\begin{figure}
\hbox to\textwidth{\hss\epsfig{file=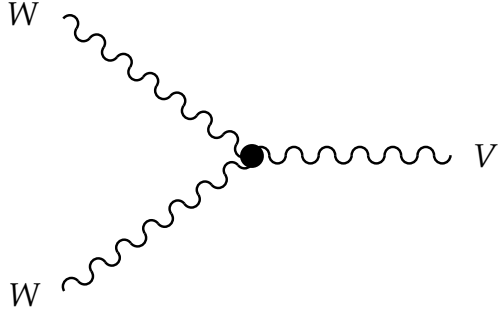,width=6.5cm}\hss}
\vskip 0.5cm
\caption{Triple-gauge-boson coupling}
\label{tribosonfig}
\end{figure}

If we restrict ourselves to C and P even couplings with electromagnetic
gauge invariance, the most general $WWV(V\equiv\gamma,Z)$
vertex(Fig.~\ref{tribosonfig}) can be parametrized
in terms of an effective Lagrangian\cite{Hagiwara}
\begin{equation}
{\cal L}^{WWV}_{eff}/g_{_{WWV}}
=ig^V_1 (W^\dagger_{\mu\nu}W^\mu V^\nu-W^{\dagger\mu}V_\nu W_{\mu\nu})
+i\kappa_{_V} W^\dagger_{\mu}W_{\nu}V^{\mu\nu}
+\frac{i\lambda_{_V}}{m^2_W} W^\dagger_{\rho\mu}{W^\mu}_{\nu}V^{\nu\rho},
\end{equation}
where $W^\mu$ and $V^\mu$ stand for the $W^-$ and the V field,
respectively, and
$W_{\mu\nu}\equiv\partial_\mu W_\nu-\partial_\nu W_\mu$,
$V_{\mu\nu}\equiv\partial_\mu V_\nu-\partial_\nu V_\mu$,
$g_{_{WW\gamma}}=-e$ and
$g_{_{WWZ}}=-e \cot\theta_{_W}$.
In SM, $g^V_1=\kappa_{_V}=1$ and $\lambda_{_V}=0$.
The static properties of the $W$,
the magnetic dipole
moment~($\mu_{_W}$) and electric quadrupole moment~($Q_{_W}$) of the
$W$  are related\cite{Aronson} to these couplings as,
\begin{equation}
\mu_{_W}=\frac{e}{2m_{_W}}(1+\kappa_\gamma+\lambda_\gamma)
\hspace{.2in}\mbox{and}\hspace{.2in}
Q_{_W}=-\frac{e}{m^2_{_W}}(\kappa_\gamma-\lambda_\gamma).
\end{equation}
Electromagnetic gauge invariance requires $g^\gamma_1=1$ and
five anomalous couplings involved in Eq.(1) survive:
$\Delta g^Z_1 \equiv g^Z_1-1,
 \Delta \kappa_\gamma \equiv \kappa_\gamma -1,
 \Delta \kappa_{_Z} \equiv \kappa_{_Z}-1,
 \lambda_\gamma$ and $\lambda_{_Z}$.
These five couplings are reduced to a smaller number by symmetry
requirements\cite{Schild}.
If we require the global $SU(2)_L$ symmetry,
then $\lambda\equiv\lambda_\gamma=\lambda_{_Z}$ and the others are zero.
Requiring an intrinsic $SU(2)_L$ symmetry, we have four independent
couplings with the relation,
\begin{equation}
1+\Delta g^1_{_Z}=-\tan^2\theta_{_W}
\frac{\Delta\kappa_\gamma}{\Delta\kappa_{_Z}}.
\end{equation}
For $WW\gamma$ coupling only, there are only two free paramters
$\lambda_\gamma$ and $\Delta\kappa_\gamma$.

To investigate the $WW\gamma$ vertex, we first review
single $W$ producton processes in $e\gamma$ and $ep$ collision.
The process $e+\gamma\rightarrow W+\nu_e$ is our main concern
in $e\gamma$ collision.
But at $ep$ collider, single $W$ production may also be possible
via the process $\gamma+q\rightarrow q^{(\prime)}+W$.
Fortunately, the cross sections of the two kinds of processes are closely
related with each other as shown below.
So we consider the two processes simultaneously.
The relevant helicity amplitude may be obtained
directly from Ref.\cite{BauerZ}. And the hard scattering cross
section of the process
$\gamma+q\rightarrow q^{(\prime)}+W$
is given in Ref.\cite{Jungil} as
\renewcommand{\theequation}{4\alph{equation}}\setcounter{equation}{0}
\begin{equation}
\left(\frac{d\hat{\sigma}} {d\hat{t}}\right)^{D}
(\gamma+q \rightarrow q^{(\prime)}+V)
=
\frac{1}{16\pi\hat{s}^{2}}
\Sigma |V|^{2},
\end{equation}
with
\begin{eqnarray}
\Sigma|V=Z|^{2}
&=&-(g_{_Z}^{2}e^{2}e_{q}^{2}g_{q}^{2})
T_{0}(\hat{u},\hat{t},\hat{s},m_{_Z}^{2})/2,\nonumber\\
\Sigma|V=W|^{2}
&=&
-(g^{2}e^{2}|V_{qq^{\prime}}|^{2})T(\hat{u},\hat{t},\hat{s},m_{_W}^{2}
,|e_q|,\kappa_\gamma,\lambda_\gamma)/2,\nonumber\\
\Sigma|V=W|^{2}_{SM}
&=&-(g^{2}e^{2}|V_{qq^\prime}|^{2})
T(\hat{u},\hat{t},\hat{s},m_{_W}^{2},|e_q|,1,0)/2\nonumber\\
&=&-(g^{2}e^{2}|V_{qq^\prime}|^{2})\left(|e_q|
-\frac{\hat{s} }{ \hat{s}+\hat{t}}\right)^{2}
T_{0}(\hat{u},\hat{t},\hat{s},m_{_W}^{2})/2,\\
 \mbox{and}
\hspace{0.4cm}g^2_q &=&\frac{1}{2}(1-4|e_q|\sin^2\theta_{_W}
+8|e_q|^{2}\sin^4\theta_{_W}),
\hspace{0.4cm}\sin^2\theta_{_W}=0.23,\nonumber
\end{eqnarray}\label{eq:dsdt}
where the subscript SM denotes the Standard Model parametrization
with $\kappa_\gamma=1$, $\lambda_\gamma=0$,
and where
\begin{eqnarray}
&T_{0}&(\hat{s},\hat{t},\hat{u},{m_{_V}}^{2})=
\left(
\frac{{\hat{t}}^{2}+{\hat{u}}^{2}+2\hat{s}m_{_V}^{2}
}{\hat{t}\hat{u}}
\right),\nonumber\\
&T&(\hat{s},\hat{t},\hat{u},m_{_W}^{2},|e_q|,\kappa_\gamma,\lambda_\gamma)
=
(|e_q|-1)^{2}\frac{\hat{u} }{ \hat{t}}
+|e_q|^{2}\frac{\hat{t} }{ \hat{u}}
+2|e_q|(|e_q|-1)m_{_W}^{2}\frac{\hat{s} }{ \hat{u}\hat{t}}
\nonumber\\
&&
-\left(
(|e_q|-1)\frac{1}{ \hat{t}}-|e_q|\frac{1}{\hat{u}}
\right)
(2\hat{s}m_{_W}^{2}-(1+\kappa_\gamma)\hat{u}\hat{t})
\frac{1}{m_{_W}^{2}-\hat{s}}
+\frac{\hat{s}}{2m_{_W}^{2}}\\
&&-\left(
2\hat{u}(\hat{u}+\hat{s})\frac{1}{m_{_W}^{2}}
+(1+\kappa_\gamma)
\left[
\hat{s}-(\hat{u}+\hat{s})^{2}\frac{1}{m_{_W}^{2}}
\right]
\right)
\frac{1}{2 (m_{_W}^{2}-\hat{s})} \nonumber
\end{eqnarray}
\newpage
\begin{eqnarray}
&&
+\left(
8{\hat{u}}^{2}-16\hat{s}m_{_W}^{2}-4(1+\kappa_\gamma){\hat{u}}^{2}
\left[1+\frac{\hat{s}}{m_{_W}^{2}}\right]\right.
\nonumber\\
&&\hspace{.7cm}
\left.+(1+\kappa_\gamma)^{2}
\left[
4\hat{u}\hat{t}+({\hat{u}}^{2}+{\hat{t}}^{2})
\frac{\hat{s}}{m_{_W}^{2}}
\right]
\right)
\frac{1}{8(m_{_W}^{2}-\hat{s})^{2}}
\nonumber\\
&&
-{\lambda_\gamma}^{2}
\frac{\hat{s}\hat{t}\hat{u}}{2m_{_W}^{4}(m_{_W}^{2}-\hat{s})}
+\lambda_\gamma(2\kappa_\gamma+\lambda_\gamma-2)\frac{\hat{s}}{8m_{_W}^{2}}
\left[
1+\frac{2\hat{t}\hat{u}}{(m_{_W}^{2}-\hat{s})^{2}}
\right].\nonumber
\end{eqnarray}
We leave the superscript $D$ in Eq.(4a) which stands
for the direct photo-process
in $ep$ collision following Ref.\cite{Jungil}.
By setting the quark charge $|e_{q}|=1$, we can obtain the matrix
elements for the processes, $e+\gamma \rightarrow \nu+W$ and
$e+\gamma \rightarrow e+Z$.
With the definitions of
$Y=\hat{s}/4m_{_W}^{2}, X=(Y-1/4)(1+\cos\hat{\theta})/2$ and
$\chi=1-\kappa_\gamma$, the differential cross section with
respect to $\hat{\theta}$, the angle between the outgoing $W$ and the
incoming photon is
\renewcommand{\theequation}{5\alph{equation}}\setcounter{equation}{0}
\begin{equation}
\frac{d\hat{\sigma}}{d\cos\hat{\theta}}
(\gamma+q \rightarrow q^{\prime}+W)=
\frac{\pi\alpha^{2}(Y-1/4)}
{128m_{_W}^{2}Y^{2}(Y-X)^{2}\sin^{2}\theta_{_W}}
F(|e_{q}|)
,
\end{equation}
where
\begin{eqnarray}
F(|e_q|)&=&
X\left[8Y-4+(8X^{2}+4X+1)/Y\right]\nonumber\\
&&-8\chi{X}(Y+X)-32\lambda_\gamma(\lambda_\gamma-\chi)YX(Y-X)
+64\lambda_\gamma^{2}YX(Y-X)^{2}\nonumber\\
&&+(\lambda_\gamma-\chi)^{2}
\left[(Y^{2}+X^{2})(4Y-4X-1)+4XY\right]\\
&&+8\xi(|e_q|)
\left[-\chi(Y+X)+(\xi(|e_q|)+2X)f\right],\nonumber
\end{eqnarray}
with\\
$$\xi(|e_q|)=(Y-X)(1-|e_q|)
\hspace{0.3cm} \mbox{and}\hspace{0.3cm}
f=\left[(Y-1/4)^{2}+(X+1/4)^{2}\right]/\left(XY\right).
\eqno(5c)
$$\\
The function $F(|e_q|=1)$ represents the matrix element for
$e+\gamma \rightarrow \nu+W$.
In the Standard Model~ {\it i.e.}~ $\lambda_\gamma=\chi=0$,
$F(|e_q|=1)$ vanishes
when the outgoing $W$ and the incoming photon are antiparallel,  $X=0$.
And that is the famous radiation zero~\cite{Mikaelian}.
It is interesting to note that the radiation zero is not a unique
feature of the Standard Model. The radiation zero will be
present~\cite{Abraham} whenever
\renewcommand{\theequation}{\arabic{equation}}\setcounter{equation}{5}
\begin{equation}
\lambda_\gamma+\kappa_\gamma=1\hspace{0.5cm}(~\mbox{or}~\lambda_\gamma=\chi~)
\hspace{0.5cm}~\mbox{and}~\hspace{0.5cm}X=0
\end{equation}
for the process $e+\gamma \rightarrow \nu+W$.

Next we focus on the total production of $W$ and $Z$ in $ep$ collisions.
In the short term these processes will be studied at
HERA($E_{e}=30~$GeV$, E_{p}=820~$GeV$, {\cal L}=200~$pb$^{-1}~$yr$^{-1}$),
while in the long term availability of
LEP~$\times$~LHC(
$E_{e}=50~$GeV$, E_{p}=8000~$GeV$, {\cal L}=1000~$pb$^{-1}~$yr$^{-1}$)
collider will give collision energies in excess of 1~TeV.
We first calculate the total cross sections for the five different processes
which contribute to single $W$ and $Z$ production at $ep$ colliders.
{}From the sum of these contributions we then calculate the ratio
$\sigma_{total}(W)/\sigma_{total}(Z)$ as a function of the anomalous
$WW\gamma$ coupling parameters $\kappa_\gamma$ and $\lambda_\gamma$.
The five processes are
\renewcommand{\theequation}{7\alph{equation}}\setcounter{equation}{0}
\begin{eqnarray}
e^{-}+p &\rightarrow& e^{-}+W^{\pm}+X,\\
&\rightarrow& \nu+W^{-}+X,\\
&\rightarrow& e^{-}+Z+X ~(Z\mbox{ from hadronic vertex}),\\
\label{eq:Zh}
&\rightarrow& e^{-}+Z+X ~(Z\mbox{ from leptonic vertex}),\\
\label{eq:Zl}
&\rightarrow& \nu+Z+X.
\end{eqnarray}

The largest contributions for $W$ and $Z$ productions come from
the processes (7a) and (7c) which are dominated by
the real photon exchange Feynman diagrams with a photon emitted from
the incoming electron,
$e^{-}+p \rightarrow \gamma_{/e}+p \rightarrow V+X$.
The dominant subprocesses for
$e+p \rightarrow V+X$ would appear to be the lowest order
$\bar{q}^{(\prime)}_{/\gamma}+q \rightarrow V$, where $q_{/\gamma}$
is a resolved quark inside the photon. However this may not be strictly true,
even at very high energies, since quarks inside the photon $q_{/\gamma}$
exist mainly through the evolution $\gamma \rightarrow q\bar{q}$.
Hence the direct process
$\gamma+q \rightarrow q^{(\prime)}+V$
could be competitive with the lowest order resolved process contribution
$\bar{q}^{(\prime)}+q \rightarrow V$.
This raises the subtle question of double counting \cite{Kim,Blumlein}.
Certain kinematic regions of the direct processes contribute to the
evolution of $q_{/\gamma}$ which is already included in the lowest
order process. Both double counting and the mass singularities are
removed \cite{Olness} if we subtract the contribution of
$\gamma+q \rightarrow q^{(\prime)}+V$
in which the $\hat{t}$-channel-exchanged quark is on-shell
and collinear with the parent photon. Thus the singularity subtracted
lowest order contribution
from the subprocesses
$\bar{q}^{(\prime)}_{/\gamma}+q\rightarrow V$ is
\renewcommand{\theequation}{8\alph{equation}}\setcounter{equation}{0}
\begin{eqnarray}
\sigma^{L}(&e^{-}&+~p \rightarrow \gamma_{/e}+p \rightarrow V+X)
=\frac{{\cal C}^{L}_{_V}}{s}\int_{m^{2}_{_V}/s}^{1}
\frac{dx_{1}}{x_{1}}\nonumber\\
&&\times\left[\sum_{qq^{\prime}}
(f_{q/e}-\tilde{f}_{q/e})(x_1,m^{2}_{_V})
f_{q^\prime /p}(\frac{m^{2}_{_V}}{x_{1}s},m^{2}_{_V})
+(q \leftrightarrow q^\prime)\right],
\end{eqnarray}
where
\begin{equation}
{\cal C}^{L}_{_W}=\frac{2{\pi}G_{_F}m^{2}_{_W}}{3\sqrt{2}}|V_{qq^\prime}|^{2},
\hspace{.5in}
{\cal C}^{L}_{_Z}=\frac{2{\pi}G_{_F}m^{2}_{_Z}}{3\sqrt{2}}g_{q}^{2}.
\end{equation}
The electron structure functions $f_{q/e}$
are obtained as usual
\renewcommand{\theequation}{\arabic{equation}}\setcounter{equation}{8}
\begin{eqnarray}
f_{q/e}(x,Q^{2})&=&
\int_{x}^{1}\frac{dy}{y}
f_{q/\gamma}(\frac{x}{y},Q^{2})f_{\gamma/e}(y),
\end{eqnarray}
where $f_{\gamma/e}$ is the appropriate
Weiz\"{a}cker-Williams approximation \cite{Weizacker} of (quasi-real)
photon radiation, and ${f}_{q/\gamma}$ is the usual photon structure
function. The part of photon structure function,
$\tilde{f}_{q/\gamma}$, which results from photon splitting at large
$x$ (with large momentum transfer), has the leading order form as
\begin{eqnarray}
\tilde{f}^{(0)}_{q/\gamma}(x,Q^{2})&=&
\frac{3\alpha{e}^{2}_{q}}{2\pi}
(1-2x+2x^{2})\log\left(\frac{Q^{2}}{\Lambda^{2}}\right),\nonumber\\
\mbox{and as before}\hspace{0.4cm}
\tilde{f}_{q/e}(x,Q^{2})&=&
\int_{x}^{1}\frac{dy}{y}
\tilde{f}^{(0)}_{q/\gamma}(\frac{x}{y},Q^{2})f_{\gamma/e}(y)\hspace{0.1cm}.
\end{eqnarray}

To obtain the total contribution from the direct subprocess,
$\gamma+q \rightarrow q^{(\prime)}+V$,
we must integrate Eq.~(4), regularizing the $\hat{t}$-pole
of the collinear singularity by cutting at the scale $\Lambda^{2}$
which determines the running of the photon structure functions
$f_{i_{/\gamma}}$. This corresponds to the subtraction used to~~redefine
the photon structure functions in Eq.~(8a). Then the hard scattering cross
sections from the direct subprocesses are
\renewcommand{\theequation}{11\alph{equation}}\setcounter{equation}{0}
\begin{equation}
\hat{\sigma}(\gamma+q \rightarrow q^{(\prime)}+V)
=\frac{{\cal C}^{D}_{_V}}{\hat{s}}\eta_{_V},
\end{equation}
where
\begin{eqnarray}
&\eta_{_{V=Z}}&(\hat{s},m^{2}_{_Z},\Lambda^{2})
=(1-2\hat{z}+2\hat{z}^{2})\log\left
(\frac{\hat{s}-m^{2}_{_Z}}{\Lambda^{2}}\right)
+\frac{1}{2}(1+2\hat{z}-3\hat{z}^{2}),\nonumber\\
&\eta_{_{V=W}}&(\hat{s},m^{2}_{_W},\Lambda^{2},|e_{q}|,\kappa_\gamma,
\lambda_\gamma)=(|e_q|-1)^{2}(1-2\hat{z}+2\hat{z}^{2})
\log
\left(
\frac{\hat{s}-m^{2}_{_W}}{\Lambda^{2}}
\right)
\nonumber\\
&&\hspace{.15in}
-\left[(1-2\hat{z}+2\hat{z}^{2})-2|e_q|(1+\kappa+2\hat{z}^{2})
+\frac{(1-\kappa)^{2}}{4\hat{z}}-\frac{(1+\kappa)^{2}}{4}
\right]\log{\hat{z}}
\nonumber\\
&&\hspace{0.15in}
+\left[\left(2\kappa+\frac{(1-\kappa)^{2}}{16}\right)\frac{1}{\hat{z}}
+\left(\frac{1}{2}+\frac{3(1+|e_q|^{2})}{2}\right)\hat{z}
\right.\nonumber\\&&\hspace{.5in}+\left.
(1+\kappa)|e_q|-\frac{(1-\kappa)^{2}}{16}+\frac{|e_q|^{2}}{2}
\right](1-\hat{z})\nonumber\\
&&\hspace{.15in}
-\frac{\lambda_\gamma^{2}}{4\hat{z}^{2}}
(\hat{z}^{2}-2\hat{z}\log{\hat{z}}-1)
\nonumber\\&&\hspace{.15in}
+\frac{\lambda_\gamma}{16\hat{z}}(2\kappa+\lambda_\gamma-2)
\left[(\hat{z}-1)(\hat{z}-9)+4(\hat{z}+1)\log{\hat{z}}\right],\\
\mbox{with}&&\nonumber\\
&&\hspace{.15in}{\cal C}^{D}_{_W}=
\frac{\alpha{G}_{_F}m^{2}_{_W}}{\sqrt{2}}|V_{qq^\prime}|^{2},
\hspace{.2in}
{\cal C}^{D}_{_Z}=\frac{\alpha{G}_{_F}m^{2}_{_Z}}{\sqrt{2}}g_{q}^{2}e_{q}^{2}
\hspace{.2in}\mbox{and}\hspace{.2in}\hat{z}=\frac{m^{2}_{_V}}{\hat{s}}.
\end{eqnarray}
The first terms in the $\eta_{_{V=W,Z}}$
represent the collinear singularity
from the $\hat{t}$-pole exchange, which is related to the photon
structure-function of Eq.~(10). This is the singularity that has already
been subtracted in Eq.~(8), and so we can now add the two contributions,
Eqs.~(8) and (12), without double counting.
The total contribution from the direct subprocess
$\gamma+q\rightarrow q^{(\prime)}+V$
is
\renewcommand{\theequation}{\arabic{equation}}\setcounter{equation}{11}
\begin{eqnarray}
\sigma^{D}(&e^{-}&+~p \rightarrow \gamma_{/e}+p \rightarrow V+X)
=\frac{{\cal C}^{D}_{_V}}{s}
\int_{m^{2}_{_V}/s}^{1}\frac{dx_{1}}{x_{1}}
\int_{m^{2}_{_V}/x_{1}s}^{1}\frac{dx_{2}}{x_{2}}
\nonumber\\
&&\times
\left[\sum_{q}
f_{\gamma/e}(x_1,Q^{2})f_{q/p}(x_{2},Q^{2})
\right]
\eta_{_V}(\hat{s}=x_{1}x_{2}s).
\end{eqnarray}

The processes (7b) and (7d),
which give a substantial
contribution as energy increases,
are dominated by configurations
where a~(quasi-real) photon is emitted~(either elastically or
quasi-elastically) from the incoming proton and subsequently scatters
off the incoming electron, {\it i.e.}
$e^{-}+p \rightarrow e^{-}+\gamma_{/p} \rightarrow
e^{-}+Z~(~$or$~ \rightarrow \nu+W^{-}).$
In these processes $Z$ is produced from leptonic vertex,
and as explained in Eq.(6) because of the famous
radiation zero, if $\lambda_\gamma+\kappa_\gamma=1$ the production of
$W^{-}$ toward the direction of incoming proton will be suppressed.
For the elastic photon, the cross section can be computed using the
electrical and magnetic form factors of the proton.
For the quasi-elastic scattering photon, the experimental information
\cite{Stein} on electromagnetic structure functions $W_{1}$ and
$W_{2}$ can be used, following Ref.~\cite{BauerV}. The hard scattering
cross section
is given
from Eqs.(5) and (11) with the obvious substitution of
$|e_{q}|=1$,
\begin{equation}
\hat{\sigma}(e^{-}+\gamma_{/p} \rightarrow
e^{-}+Z~\mbox{~or~}~\nu+W^{-})
=\frac{{\cal C}^{D}_{_{V=W,Z}}}{\hat{s}}
\eta_{_{V=W,Z}}(|e_{q}|=1).
\end{equation}
Notice that since in these processes there is no contribution from
$\hat t$--pole quark exchange diagram, which dominates for the
processes (7a) and (7c), the production cross section of
$e p \rightarrow \nu W^- X$ is significantly smaller compared to
$e p \rightarrow e W^{\pm} X$. However, due to the contribution from
the diagram with $WW\gamma$ vertex the rate for $e p \rightarrow \nu W^- X$
grows more rapidly with energy than the rate for $e p \rightarrow e Z X$,
as shown in  Table 1.
For process (7e), which is a pure charged current process,
we simply
use the results of Bauer {\it et. al.} \cite{BauerV} to add to the
contributions from (7c) and (7d). The contribution from this process
to the total $Z$ production cross section is almost negligible even at
LEP~$\times$~LHC $ep$ collider energies, as
can be seen in Table 2.

Finally, as explained before, we again emphasize that for the
processes of $e \gamma$ collisions, $e + \gamma \rightarrow W^- + \nu$,
and $e + \gamma \rightarrow Z + e$, we can get all the relavant
results by setting the quark charge $|e_q|=1$.
\section{The Process in $e\gamma$ Collisions}
Let us now consider the $W$ decay to final state fermions in
$e\gamma$ collisions.
The net process is represented by $e\gamma\rightarrow \nu_e f\bar{f}$.
In the hadronic decay of the $W$ boson,
$e+\gamma\rightarrow \nu_e+q+\bar{q}$,
there are two hadronic jets and large missing transverse
momentum($\not{p_{_T}}$) due to the neutrino from the initial electron
beam.
But in leptonic decay,
$e+\gamma\rightarrow \nu_e+\mu+\bar{\nu}_\mu$,
we can detect only $\mu$ with the missing $p_{_T}$
which is attributed to the neutrinos.
If we consider single $\mu$ producton in $e\gamma$ collision,
it must be produced via the process
$e+\gamma\rightarrow \nu_e+\mu+\bar{\nu}_\mu$.
In this respect, the single $\mu$ production in $e\gamma$ collision has
strong merit for being studied.
Single $\mu$ production in $e\gamma$ collisions
has recently  been studied in Ref.\cite{Yeh,step}.
But they restricted the production of $\mu$ as a decay product of real $W$
at a future TeV energy colliders.
Here we include the single $\mu$ production via virtual $W$.

\begin{figure}
\hbox to\textwidth{\hss\epsfig{file=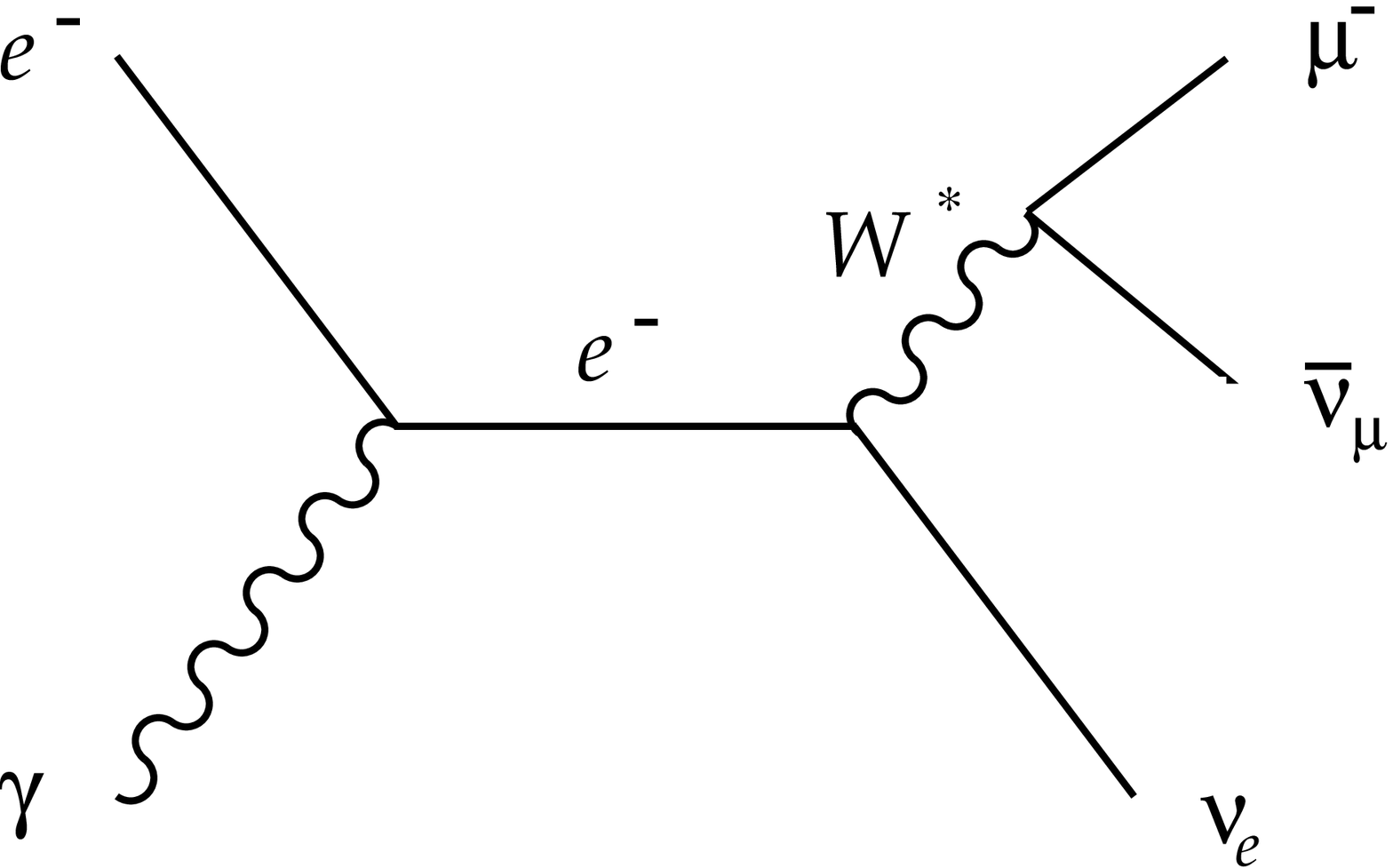,width=6.5cm}\hss}
\hbox to\textwidth{\hss\epsfig{file=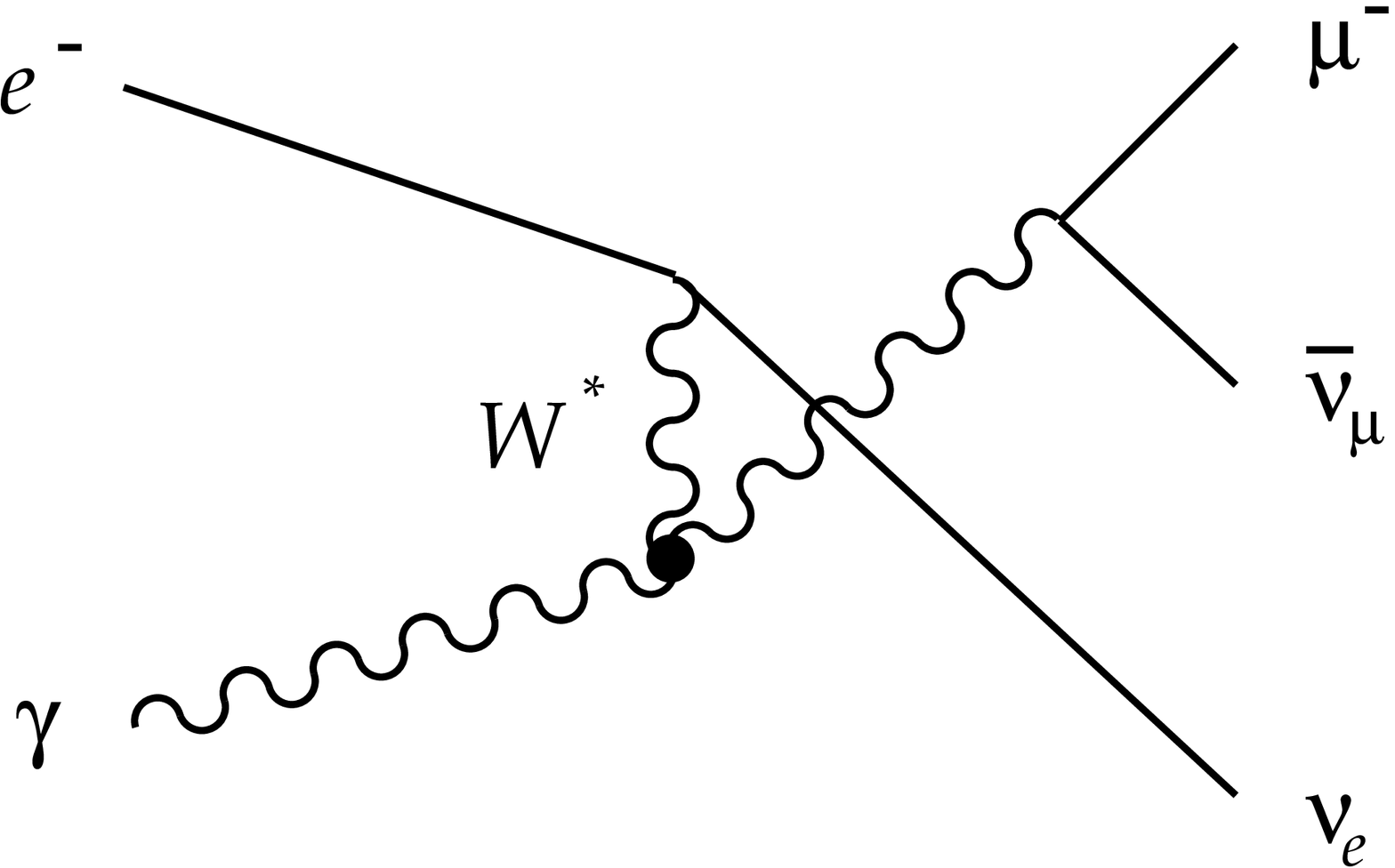,width=6.5cm}
                   \hss\epsfig{file=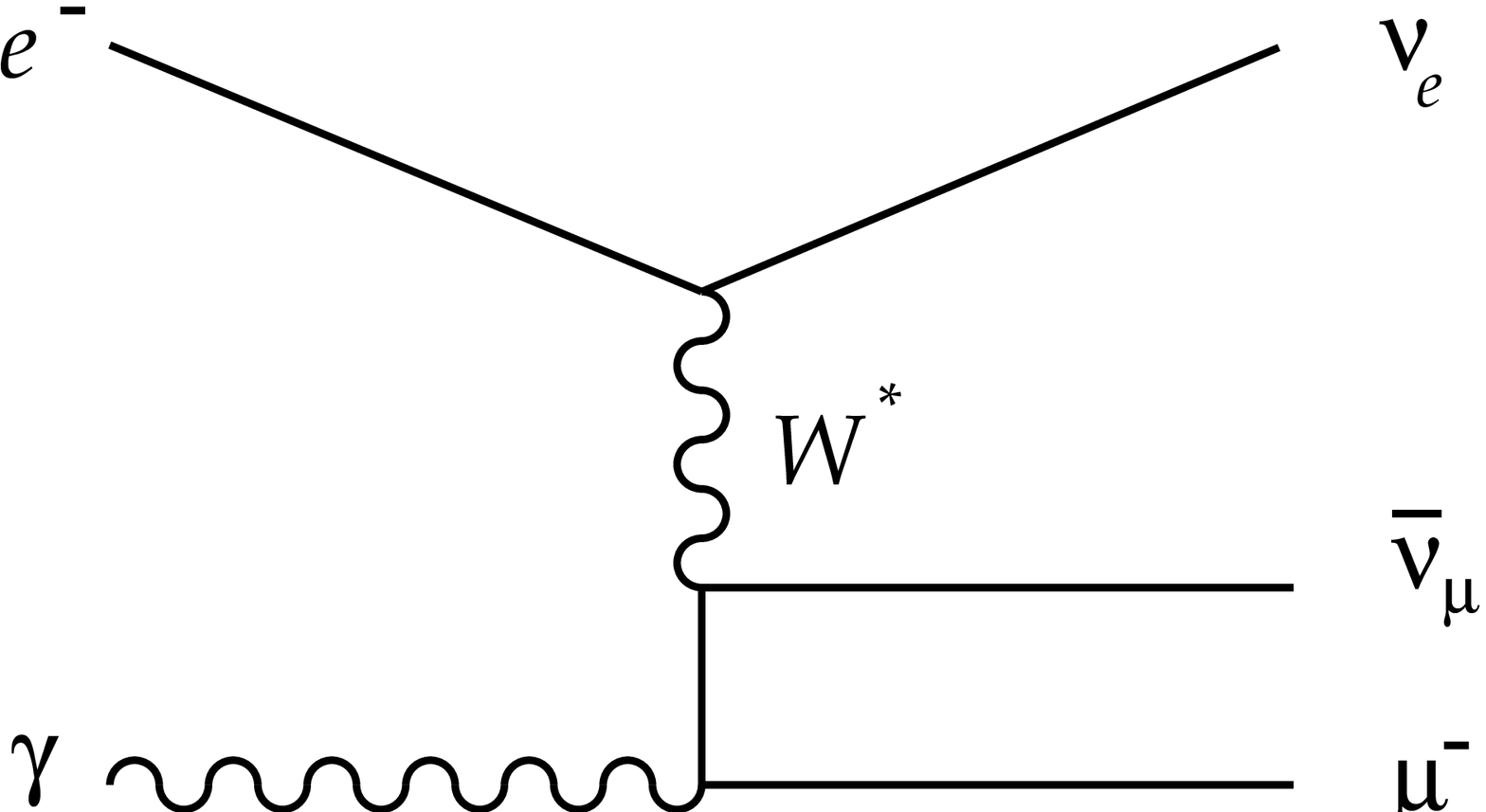,width=6.5cm}\hss}
\vskip 0.5cm
\caption{Feynmean diagrams for  the process
$e+\gamma\rightarrow \nu_e+\mu+\bar{\nu}_\mu$}
\label{egmfig}
\end{figure}

The lowest order tree level Feynman diagrams are given in Fig.
\ref{egmfig}.
Let us introduce invariant variables commonly used for
2$\rightarrow$3 processes as,
\begin{eqnarray}
s&=&(e+\gamma\hskip .1cm)^2 ,\hskip 1cm
s^\prime\hskip .1cm=\hskip .1cm(\nu_e  +\mu )^2,
\nonumber\\
t&=&(e-\nu_e )^2 , \hskip 1cm
t^\prime\hskip .1cm=\hskip .1cm(\gamma -\nu_e)^2,
\\
u&=&(e-\mu   \hskip .1cm)^2 ,\hskip 1cm
u^\prime\hskip .1cm=\hskip .1cm(\gamma -\mu \hskip .1cm )^2,\nonumber
\end{eqnarray}
where we express  the momentum of each particle by its name.

The matrix element ${\cal M}$ is given by,
\begin{equation}
{\cal M}=
i \frac{g^2e}{2}
\left({\cal M}_a+{\cal M}_b+{\cal M}_c\right),
\end{equation}
where
\begin{eqnarray}
{\cal M}_a&=&
\bar{u}_L(\nu_e)
\gamma^\alpha \frac{\not{e}+\not{\gamma}}{s}\not{\epsilon}_\gamma
u_L(e)
\frac{-g_{\alpha\beta}} {s+t+t^\prime-m_{_W}^2}
\bar{u}_L(\mu)
\gamma^\beta
v_L(\bar{\nu}_\mu),\nonumber\\
{\cal M}_b&=&
\bar{u}_L(\nu_e)
\gamma^\alpha
u_L(e)
\frac{\Gamma_{\alpha\beta_\sigma} \epsilon_\gamma^\sigma}
     {(t-m_{_W}^2)(s+t+t^\prime-m_{_W}^2)}
\bar{u}_L(\mu)
\gamma^\beta
v_L(\bar{\nu}_\mu),
\\
{\cal M}_c&=&
\bar{u}_L(\nu_e)
\gamma^\alpha
u_L(e)
\frac{-g_{\alpha\beta}}{t-m_{_W}^2}
\bar{u}_L(\mu)
\not{\epsilon}_\gamma
\frac{\not\mu-\not\gamma}{u^\prime}
\gamma^\beta
v_L(\bar{\nu}\mu).\nonumber
\end{eqnarray}
Here the $WW\gamma$ vertex factor
$\Gamma_{\mu\nu\sigma}$ in SM is defined by
\begin{equation}
\Gamma_{\mu\nu\sigma}=
 (k_+ - k_-     )_\sigma g_{\mu\nu}
+(k_- - k_\gamma)_\mu    g_{\nu\sigma}
+(k_\gamma-k_+  )_\nu    g_{\mu\sigma},
\end{equation}
where $k_+$, $k_-$, $k_\gamma$
,$\mu$, $\nu$ and $\sigma$
are the momentums with incoming direction and corresponding indices
of $W^+$, $W^-$, and $\gamma$, respectively.

We present averaged amplitude squared for the process
$e+\gamma\rightarrow \nu_e+\mu+\bar{\nu}_\mu$ in compactly factorized form
{}.
We get the spin averaged amplitude squared as,
\begin{eqnarray}
|{\cal M}|^2&=&
g^4 e^2
\frac{(s+t+u)^2 +(s+t^\prime+u^\prime)^2 +(t+u)^2
     +2(t+u)(s+t^\prime+u^\prime)}
{2 (t-m_{_W}^2) (s+t+t^\prime-m_{_W}^2) }
\nonumber\\
&&\hskip .2cm
\times\left[
                \frac{m_{_W}^2}{u^\prime (t-m_{_W}^2)}
              + \frac{m_{_W}^2}{s+t+t^\prime-m_{_W}^2}
                               (\frac{1}{s}+\frac{1}{t-m_{_W}^2})
              -\frac{u}{s u^\prime}
        \right].
\end{eqnarray}

The simplified amplitude squared as a general function of $\kappa_\gamma$
and $\lambda_\gamma$ will be presented elsewhere.
Here we give the result for the case of the SM only.

\section{Numerical Results and Discussions}

In Table 1 we show the total $W^{\pm}$ production cross section at HERA
and LEP~$\times$~LHC $ep$ colliders for a range of values of
the anomalous $WW\gamma$
coupling parameters $\kappa_\gamma$ and $\lambda_\gamma$.
The error range represents
the variation in the cross section by varying the theoretical input
parameters as follows~:~$m^{2}_{_V}/10\leq{Q}^{2}\leq{m}^{2}_{_V}$
and $Q^2=p^2_{_T}(V)$, photon
structure functions $f_{q/\gamma}$ from DG~\cite{Drees} and
DO$+$VMD~\cite{Duke}, and proton structure functions $f_{q/p}$
from EHLQ1~\cite{Eichten}, HMRS(B)~\cite{Harriman} and
GRV~\cite{Gluck}.
There are very strong $Q^2$ dependences in the total cross sections
of $W,Z$ production. We find that for any values of $\kappa_\gamma$ and
$\lambda_\gamma$,
always $\sigma_{_{W,Z}}(Q^2=p^2_{_T})\leq\sigma_{_{W,Z}}(Q^2=m^2_{_V}/10)
\leq\sigma_{_{W,Z}}(Q^2=m^2_{_V})$.
We also find that there exists a quite strong dependence on the structure
functions of  $f_{q_{/p}}$ and $f_{q_{/\gamma}}$ for $W^+$ production,
but almost no
dependence for $W^-, Z$ production at HERA energies. Fortunately the ratio
of $W$ and $Z$ production cross sections give much weaker dependence
on the variation of the theoretical input parameters, as can be seen
in Tables 1,~2 and 3.


\begin{table}
\begin{center}
\begin{tabular}{|c|c|c|c|}
\multicolumn{4}{c}{\bf HERA $W$-production Cross-section~(in pb)}
\\ \hline
&$ep\rightarrow{W^+}X$& $ep\rightarrow{W^-}X$& $ep\rightarrow{W^\pm}X$\\ \hline
\makebox[1.4in]{$\lambda_\gamma=0$, $\kappa_\gamma=0.0$}
&\makebox[1.25in]{0.46 $\pm$ 0.04}
&\makebox[1.25in]{0.56 $\pm$ 0.03}
&\makebox[1.25in]{1.02 $\pm$ 0.07}\\\hline
$\lambda_\gamma=0$, $\kappa_\gamma=0.5$&0.53 $\pm$ 0.04&0.61 $\pm$
0.04&1.14 $\pm$ 0.07
\\\hline
$\lambda_\gamma=0$, $\kappa_\gamma=1.0$&0.63 $\pm$ 0.05&0.69 $\pm$
0.03&1.31 $\pm$ 0.07
\\\hline
$\lambda_\gamma=0$, $\kappa_\gamma=1.5$&0.75 $\pm$ 0.03&0.79 $\pm$
0.03&1.54 $\pm$ 0.06
\\\hline
$\lambda_\gamma=0$, $\kappa_\gamma=2.0$&0.92 $\pm$ 0.06&0.94 $\pm$
0.03&1.85 $\pm$ 0.08
\\\hline
$\lambda_\gamma=0.0$, $\kappa_\gamma=1$&0.63 $\pm$ 0.05&0.69 $\pm$
0.03&1.31 $\pm$ 0.07
\\\hline
$\lambda_\gamma=0.5$, $\kappa_\gamma=1$&0.63 $\pm$ 0.04&0.71 $\pm$
0.03&1.33 $\pm$ 0.06
\\\hline
$\lambda_\gamma=1.0$, $\kappa_\gamma=1$&0.67 $\pm$ 0.03&0.72 $\pm$
0.03&1.39 $\pm$ 0.05
\\\hline
$\lambda_\gamma=1.5$, $\kappa_\gamma=1$&0.71 $\pm$ 0.04&0.77 $\pm$
0.04&1.48 $\pm$ 0.07
\\\hline
$\lambda_\gamma=2.0$, $\kappa_\gamma=1$&0.77 $\pm$ 0.04&0.83 $\pm$
0.03&1.61 $\pm$ 0.07
\\\hline
\multicolumn{4}{c}{}\\
\multicolumn{4}{c}{\bf LEP$\times$LHC $W$-production Cross-section~(in pb)}
\\ \hline
&$ep\rightarrow{W^+}X$& $ep\rightarrow{W^-}X$& $ep\rightarrow{W^\pm}X$
\\ \hline
\makebox[1.4in]{$\lambda_\gamma=0$, $\kappa_\gamma=0.0$}
&\makebox[1.25in]{6.17 $\pm$ 1.17}
&\makebox[1.25in]{7.38 $\pm$ 1.27}
&\makebox[1.25in]{13.63 $\pm$ 2.38}
\\\hline
$\lambda_\gamma=0$, $\kappa_\gamma=0.5$&7.64 $\pm$ 1.29&8.82 $\pm$
1.35&16.34 $\pm$ 2.48
\\\hline
$\lambda_\gamma=0$, $\kappa_\gamma=1.0$&9.78 $\pm$ 1.37&11.49 $\pm$
1.54&21.16 $\pm$ 2.73
\\\hline
$\lambda_\gamma=0$, $\kappa_\gamma=1.5$&13.12 $\pm$ 1.64&15.56 $\pm$
1.54&28.77 $\pm$ 2.65
\\\hline
$\lambda_\gamma=0$, $\kappa_\gamma=2.0$&17.63 $\pm$ 1.58&20.95 $\pm$
1.23&38.63 $\pm$ 2.76
\\\hline
$\lambda_\gamma=0.0$, $\kappa_\gamma=1$&9.78 $\pm$ 1.37&11.49 $\pm$
1.54&21.16 $\pm$ 2.73
\\\hline
$\lambda_\gamma=0.5$, $\kappa_\gamma=1$&11.56 $\pm$ 1.44&13.49 $\pm$
1.33&25.09 $\pm$ 2.67
\\\hline
$\lambda_\gamma=1.0$, $\kappa_\gamma=1$&16.43 $\pm$ 1.67&19.84 $\pm$
1.35&36.17 $\pm$ 2.82
\\\hline
$\lambda_\gamma=1.5$, $\kappa_\gamma=1$&24.30 $\pm$ 1.74&31.16 $\pm$
2.44&54.77 $\pm$ 3.13
\\\hline
$\lambda_\gamma=2.0$, $\kappa_\gamma=1$&35.69 $\pm$ 2.36&44.61 $\pm$
2.54&79.68 $\pm$ 4.27
\\\hline
\end{tabular}
\end{center}
\caption{Total $W$-production cross sections (in pb) at HERA and at
LEP~$\times$~LHC, as a function of anomalous $WW\gamma$ coupling
parameters $\kappa_\gamma$ and $\lambda_\gamma$. The error range
represents the
uncertainties in the cross sections by varying the theoretical input
parameters : $m^{2}_{V}/10 \leq Q^{2} \leq m^{2}_{V}$,
photon structure functions $f_{q_{/\gamma}}$~(DG and DO+VMD),
and proton structure functions $f_{q_{/p}}$
(EHLQ1, HMRS(B) and GRV).}
\end{table}
\begin{table}
\begin{center}
\begin{tabular}{|c|l|c|c|}
\multicolumn{4}{c}{\bf $W$ and $Z$ production Cross-sections~(in pb)}
\\ \hline
\multicolumn{1}{|c|}{Eq.}
&\hspace{1.2cm}\makebox[4.0cm][l]{Process}
&HERA           &LEP$\times$LHC \\ \hline
\makebox[1.2cm]{(7a)}
&\hspace{0.7cm}\makebox[4.5cm][l] {$ep$ ${\rightarrow}$ $eW^{+}X$ }
&\makebox[3.6cm]{0.63 $\pm$ 0.05}
&\makebox[3.6cm]{9.78 $\pm$ 1.37}\\
(7a)
&\hspace{0.7cm}$ep$ ${\rightarrow}$ $eW^{-}X$
&0.63 $\pm$ 0.03&8.80 $\pm$ 1.54\\
(7b)
&\hspace{0.7cm}$ep$ ${\rightarrow}$ ${\nu}W^{-}X$
&0.06&2.69 \\ \hline
&\hspace{0.7cm}$ep$ ${\rightarrow}$ $W^{\pm}X$
&1.31 $\pm$ 0.07&21.16 $\pm$ 2.73 \\ \hline
(7c)
&\hspace{0.7cm}$ep$ ${\rightarrow}$ $eZX$ (hadronic)
&0.31 $\pm$ 0.01&2.58 $\pm$ 0.75 \\
(7d)
&\hspace{0.7cm}$ep$ ${\rightarrow}$ $eZX$ (leptonic)  &0.16&1.17\\
(7e)
&\hspace{0.7cm}$ep$ ${\rightarrow}$ ${\nu}ZX$        &0.004&0.61\\
\hline
&\hspace{0.7cm}$ep$ ${\rightarrow}$ $ZX$
&0.47 $\pm$ 0.01  &4.36 $\pm$ 0.75\\ \hline
\end{tabular}
\end{center}
\caption{
The cross sections~(in pb) for the various $W,~Z$ production
channels  at HERA and LEP~$\times$~LHC $ep$ colliders.
The errors represent the variation in cross~~ sections obtained by
varying the theoretical input parameters:
$m^{2}_{_V}/10 \leq Q^{2} \leq m^{2}_{_V}$ and
$Q^2=p^2_{_T}(V)$,
photon structure functions $f_{q/\gamma}$~(DG
and DO+VMD),
and proton structure functions $f_{q/p}$~(
EHLQ1, HMRS(B)
and GRV). For $W$ production,
$\kappa_\gamma=1$ and $\lambda_\gamma=0$ are assumed.
}
\end{table}
\begin{table}
\begin{center}
\begin{tabular}{|c|c|c|c|}
\multicolumn{4}{c}{\bf HERA $W/Z$-production Ratio} \\ \hline
&{$\sigma(W^+)/\sigma(Z)$}
&{$\sigma(W^-)/\sigma(Z)$}
& {$\sigma(W^\pm)/\sigma(Z)$}\\ \hline
\makebox[1.4in]{$\lambda_\gamma=0$, $\kappa_\gamma=0.0$}
&\makebox[1.25in]{0.98 $\pm$ 0.09}
&\makebox[1.25in]{1.20 $\pm$ 0.07}
&\makebox[1.25in]{2.18 $\pm$ 0.14}\\\hline
$\lambda_\gamma=0$, $\kappa_\gamma=0.5$&1.12 $\pm$ 0.07&1.30 $\pm$
0.09&2.41 $\pm$ 0.16
\\\hline
$\lambda_\gamma=0$, $\kappa_\gamma=1.0$&1.31 $\pm$ 0.08&1.45 $\pm$
0.05&2.76 $\pm$ 0.12
\\\hline
$\lambda_\gamma=0$, $\kappa_\gamma=1.5$&1.58 $\pm$ 0.08&1.67 $\pm$
0.09&3.24 $\pm$ 0.16
\\\hline
$\lambda_\gamma=0$, $\kappa_\gamma=2.0$&1.95 $\pm$ 0.14&1.97 $\pm$
0.09&3.91 $\pm$ 0.22
\\\hline
$\lambda_\gamma=0.0$, $\kappa_\gamma=1$&1.31 $\pm$ 0.08&1.45 $\pm$
0.05&2.76 $\pm$ 0.12
\\\hline
$\lambda_\gamma=0.5$, $\kappa_\gamma=1$&1.33 $\pm$ 0.08&1.49 $\pm$
0.09&2.81 $\pm$ 0.15
\\\hline
$\lambda_\gamma=1.0$, $\kappa_\gamma=1$&1.40 $\pm$ 0.09&1.53 $\pm$
0.06&2.93 $\pm$ 0.14
\\\hline
$\lambda_\gamma=1.5$, $\kappa_\gamma=1$&1.51 $\pm$ 0.09&1.63 $\pm$
0.07&3.14 $\pm$ 0.16
\\\hline
$\lambda_\gamma=2.0$, $\kappa_\gamma=1$&1.65 $\pm$ 0.10&1.76 $\pm$
0.07&3.40 $\pm$ 0.15
\\\hline
\multicolumn{4}{c}{}\\
\multicolumn{4}{c}{\bf LEP$\times$LHC $W/Z$-production Ratio} \\ \hline
&{$\sigma(W^+)/\sigma(Z)$}
&{$\sigma(W^-)/\sigma(Z)$}
&{$\sigma(W^\pm)/\sigma(Z)$}\\ \hline
\makebox[1.4in]{$\lambda=0$, $\kappa=0.0$}
&\makebox[1.25in]{1.40 $\pm$ 0.14}
&\makebox[1.25in]{1.69 $\pm$ 0.11}
&\makebox[1.25in]{3.10 $\pm$ 0.24}\\\hline
$\lambda_\gamma=0$, $\kappa_\gamma=0.5$&1.69 $\pm$ 0.11&1.98 $\pm$
0.14&3.68 $\pm$ 0.23
\\\hline
$\lambda_\gamma=0$, $\kappa_\gamma=1.0$&2.14 $\pm$ 0.12&2.54 $\pm$
0.11&4.66 $\pm$ 0.22
\\\hline
$\lambda_\gamma=0$, $\kappa_\gamma=1.5$&2.86 $\pm$ 0.30&3.44 $\pm$
0.26&6.35 $\pm$ 0.47
\\\hline
$\lambda_\gamma=0$, $\kappa_\gamma=2.0$&3.89 $\pm$ 0.31&4.75 $\pm$
0.42&8.60 $\pm$ 0.68
\\\hline
$\lambda_\gamma=0.0$, $\kappa_\gamma=1$&2.14 $\pm$ 0.12&2.54 $\pm$
0.11&4.66 $\pm$ 0.22
\\\hline
$\lambda_\gamma=0.5$, $\kappa_\gamma=1$&2.57 $\pm$ 0.13&2.97 $\pm$
0.22&5.55 $\pm$ 0.33
\\\hline
$\lambda_\gamma=1.0$, $\kappa_\gamma=1$&3.68 $\pm$ 0.28&4.52 $\pm$
0.36&8.21 $\pm$ 0.59
\\\hline
$\lambda_\gamma=1.5$, $\kappa_\gamma=1$&5.42 $\pm$ 0.66&6.87 $\pm$
0.69&12.33 $\pm$ 1.31
\\\hline
$\lambda_\gamma=2.0$, $\kappa_\gamma=1$&8.09 $\pm$ 0.94&10.25 $\pm$
1.43&18.32 $\pm$ 2.28
\\\hline
\end{tabular}
\end{center}
\caption{
Production cross section ratio of $W/Z$ as a function of $\kappa_\gamma$ and
$\lambda_\gamma$ at HERA and at LEP~$\times$~LHC. We first set
$\lambda_\gamma$ to
its Standard Model values~($\lambda_\gamma=0$) and then vary $\lambda_\gamma$
and vice versa.}
\label{wzratiotable}
\end{table}
It is quite important to note that
once photoproduction experiments at
HERA determine $f_{q/p}$ and $f_{q/\gamma}$ more precisely,
we will be able to predict the total cross sections for each process
with much greater accuracy.
The subtraction terms $\tilde{f}_{q/\gamma}$ of Eq.~(8) have been here
calculated using the leading order photon splitting function
as in Eq.~(10), the same prescription also used in Ref.~[19].
In our previous study [6], cut-off dependent higher
order terms were included in $\tilde{f}_{q/\gamma}$ to calculate
the processes (7a) and (7c).

We show in Table 2 the cross sections
for the various $W$ and $Z$ production channels
at the HERA and LEP~$\times$~LHC $ep$ colliders. The errors represent
the variation in cross sections obtained by varying the input
parameters, as in Table 1.
We find that our results of Table 2 agree quite well with the results
of Ref. [19],  Table 5.
Here we note several comments for Table 2;
(i) $W$ production cross sections are with the Standard Model
parametrization, {\it i.e.} $\kappa_\gamma=1$ and $\lambda_\gamma=0$.
(ii) Notice that the importance of $Z$ production from the
leptonic vertex, (7d).
(iii) As explained earlier, due to the contribution from the
diagram with $WW\gamma$ vertex, the rate for $\nu W^- X$ production
grows rapidly with energy.
(iv) We have not included the contribution from $W$ and $Z$ exchange
diagrams, which is very small at HERA energies [19].

With the anticipated luminocities of
${\cal L}=200 $ pb $^{-1}$ yr $^{-1}$ (HERA) and
${\cal L}=1000 $ pb $^{-1}$ yr $^{-1}$ (LEP $\times$ LHC),
the total $Z$ production cross section corresponds to
84~events/yr~(HERA) and 5400 events/yr (LEP $\times$ LHC).
After including a $6.7~\%$ leptonic branching
ratio~({\it i.e.}~$Z \rightarrow e^{+}e^{-},\mu^{+}\mu^{-}$),
the event numbers become about 6~events/yr (HERA) and
360~events/yr (LEP~$\times$~LHC).

In Table 3 we show the ratio
$\sigma(W^{+})/\sigma(Z)$, $\sigma(W^{-})/\sigma(Z)$ and
$\sigma(W^{+}+W^{-})/\sigma(Z)$ for the various values of $\kappa_\gamma$
and $\lambda_\gamma$. The input parameters have beeen varied as in Table~1.
Note also that we have not included the uncertainties due to higher
order perturbative QCD corrections. While these are expected to have
non-negligible effect on the absolute $W$ and $Z$
cross sections - as in $pp$ and $p\bar{p}$ collisions -
the ratio of $W$ to $Z$ cross sections is one of the most reliable
predictions of QCD, as every diagram, except for the diagrams
with $WW\gamma$ vertex, producing a $W$ also produces $Z$
up to $O(\alpha^{2}_{s})$ where additional diagrams produce $Z$
via a triangular quark loop\cite{tri}.
Even this contribution would vanish for equal-mass up-~and down-type
quarks.
These $O(\alpha^{2}_{s})$ contributions have also been calculated\cite{dis},
and are less than $1\%$ in $pp(\bar{p})$ colliders even for a very
heavy top quark. Henceforth we ignore higher order QCD corrections,
and investigate the uncertainties due to the theoretical input
parameters as explained.

To obtain an experimentally measurable ratio
$\sigma(ep \rightarrow W^{\pm} \rightarrow l\nu)/
\sigma(ep \rightarrow Z \rightarrow l^{+}l^{-})$
we must multiply the cross section ratio
$\sigma(W)/\sigma(Z)$ by the leptonic branching ratio factor
\begin{equation}
R_{BR}(m_{t}>m_{W}-m_{b},N_\nu=3)\equiv
\frac{BR(W^{\pm} \rightarrow l\nu)}{BR(Z^{\pm} \rightarrow l^{+}l^{-})}
=3.23.
\end{equation}
After 5 years of running, HERA will produce about 30 events of
$e+p \rightarrow Z+X \rightarrow l^{+}+l^{-}+X$,
and this will enable us to determine $\kappa_\gamma$ and $\lambda_\gamma$
with a precision of order
\begin{eqnarray}
\Delta\kappa_\gamma&\approx&{\pm}0.3\hspace{.1in}
\mbox{for}\hspace{.1in}\lambda_\gamma=0,
\nonumber\\
\Delta\lambda_\gamma&\approx&{\pm}0.8\hspace{.1in}
\mbox{for}\hspace{.1in}\kappa_\gamma=1,
\end{eqnarray}
which are comparable with the expected constraints from
the future LEP-II ~$e^{+}e^{-}$~ experiment.
At LEP~$\times$~LHC, one year's running will give
\begin{eqnarray}
\Delta\kappa_\gamma&\approx&{\pm}0.2\hspace{.1in}
\mbox{for}\hspace{.1in}\lambda_\gamma=0,
\nonumber\\
\Delta\lambda_\gamma&\approx&{\pm}0.3\hspace{.1in}
\mbox{for}\hspace{.1in}\kappa_\gamma=1.
\end{eqnarray}
\section{Conclusion}

We have shown how measurements of weak boson production at high energy
electron-proton and electron-photon colliders
can provide important information on anomalous
$WW\gamma$ couplings.

In $ep$ collisions, we have analyzed the production of massive gauge
bosons -- $W$ and $Z$. We have included both direct and indirect processes,
involing the parton structure of the photon, taking careful account of the
double counting problem for the latter.
We have also argued that the ratio of $W$ and $Z$ production cross sections
is particularly suited to an experimental determination of the anomalous
$WW\gamma$ coupling parameters $\kappa_\gamma$ and $\lambda_\gamma$, being
relatively
insensitive to uncertainties in the theoretical input parameters.
In fact, with more precise measurements of the input parameters in the next
few years - in particular the photon structure functions - the errors in the
measured $\kappa_\gamma$ and $\lambda_\gamma$ values will ultimately be
obtained by
the statistical error from the small number of $Z$ events at HERA.
In this respect, the higher energy LEP~$\times$~LHC collider offers a
significant improvement. Finally we note that our estimated precision on
$\kappa_\gamma$ and $\lambda_\gamma$ for both $ep$ colliders,
Eqs.~(20) and (21), is an order of magnitude greater than existing
measurements from $W\gamma$ production
at $p\bar{p}$ collider~\cite{Alitti}.

The Pohang Light Source~(PLS) facility can produce a high flux
($10^7$/sec) of background-free $\gamma$-rays after installation of a laser
backscattering system. The maximum energy of the $\gamma$-rays will reach
300 MeV when the facility is upgraded to an electron energy of 2.5 GeV.
In a later stage of the project, a Free Electron Laser can be
expected to extend the $\gamma$-ray energy up to 1 GeV\cite{pls}.
If we consider the $e\gamma$ collision
at PLS($E_e\sim$ 2.5GeV,$E_\gamma\sim$ 1 GeV),
the total cross section of the process
$e+\gamma\rightarrow W^*(\rightarrow+\mu+\bar{\nu}_\mu)+\nu_e$
is about $10^{-4}$pb. To get 1 event/yr, very large luminosity
must be obtained(${\cal L}\sim 10^{33}$/cm$^2$/sec).
Therefore, this process is not adequate to be detected at PLS
for the time being.

Attempts are at present under way by many authors to constrain the
parameter space of $\lambda_\gamma$ and $\kappa_\gamma$ by considering
various experimental results; production of $W+\gamma$ at $p\bar{p}$
collider\cite{berger}, process $e\gamma\rightarrow W\nu$
at future $e^+e^-$ and $e\gamma$ colliders\cite{Abraham,ee}, and also from
present low energy data\cite{low}.We are now studying this process at HERA via
$e+\gamma_{/p}\rightarrow\mu+\bar{\nu}_\mu+\nu_e$
considering both $\kappa_\gamma$ and $\lambda_\gamma$.
And those approaches should be regarded as complementary
in the efforts to find new physics beyond the Standard Model.
\begin{center}{\bf Acknowledgements}\end{center}
The work of CSK was supported
in part by the Korean Science and Engineering  Foundation,
in part by Non-Direct-Research-Fund, Korea Research Foundation 1993,
in part by the Center for Theoretical Physics, Seoul
National University, in part by Yonsei University Faculty Research
Grant, and in part by the Basic Science Research Institute Program,
Ministry of Education, 1994,  Project No. BSRI-94-2425.

\end{document}